\documentstyle[prd,aps,psfig,axodraw]{revtex}
\newcommand{\SI}{\int\hspace{-0.52cm}\sum_{\{K\}}}

\begin{document}

\title{\Large{Chiral  Susceptibility in Hard Thermal Loop
Approximation}}

\author{Purnendu Chakraborty$^1$, Munshi G. Mustafa$^{1,2}$\thanks{On leave of
absence from Saha Institute of Nuclear Physics.} and Markus H. Thoma$^3$}

\address{$^1$Theory Group, Saha Institute of Nuclear Physics, 1/AF Bidhan Nagar,
Kolkata 700 064, India}

\address{$^2$ Institut f\"ur Theoretische Physik, Universit\"at Giessen,
35392 Giessen, Germany}


\address{$^3$Centre for Interdisciplinary Plasma Science, 
Max-Planck-Institut f\"ur extraterrestrische Physik,
P.O. Box 1312, 85741 Garching, Germany}

\maketitle

\vspace{0.4in}

\begin{abstract}
The static and dynamic chiral susceptibilities in the quark-gluon plasma 
are calculated within the lowest order perturbative QCD at finite temperature and the
Hard Thermal Loop resummation technique using an effective quark propagator. 
After regularization of ultraviolet divergences, the Hard Thermal Loop
results are compared to QCD lattice simulations.
\end{abstract}

\vspace{0.4in}

\section{Introduction}

The main aim of relativistic heavy-ion collisions is to unravel 
the basic properties of QCD associated with, e.g., its phase diagram. Two
of the main features characterizing the ground state of the theory,
confinement and the spontaneous breaking of chiral symmetry, are expected
to cease in a possibly common phase transition at finite temperature and/or
density~\cite{satz}. The dynamical breaking of chiral symmetry is associated
with the condensation of quark-antiquark pairs in the QCD vacuum.
As the temperature and/or baryon density increase, 
the QCD vacuum undergoes a phase transition to the chirally symmetric
phase where the quark-antiquark condensate, the order 
parameter of the chiral phase transition, vanishes. 
The fluctuation of this order
parameter is related to the associated susceptibilities~\cite{hatsuda}. 
The quantity of interest here is the chiral/scalar density susceptibility 
which measures the response of the chiral condensate/scalar density to
the variation of the current quark mass. The chiral susceptibility was 
recently measured in lattice QCD~\cite{satz,laer} with two light flavors. 
The chiral susceptibility
has also been studied in chiral perturbation theory \cite{smilga},
in the multiflavor Schwinger model for a 
small nonzero quark mass~\cite{smilga}, in the NJL model \cite{zhuang}, and
using the Dyson-Schwinger equation \cite{blaschke}.
In this paper, we will investigate the chiral/scalar density susceptibility 
following the same line as in the case of the quark number susceptibility \cite{mun}
and the free energy of the quark-gluon plasma~\cite{Andersen99,andersen00} within the Hard 
Thermal Loop (HTL) approximation~\cite{htl}, which selectively resums higher order 
corrections corresponding to medium effects such as 
screening, quasiparticles masses and Landau damping beyond the usual
perturbation theory.  
  
\section{Definition}

\subsection{General}

Let \( {\cal O}_{\alpha } \) be a Heisenberg operator. In a static and uniform
external field \( {\cal F}_{\alpha } \), the (induced) 
expectation value of the operator \( {\cal O}_\alpha \left( 0,\vec{x}
\right) \) is written~\cite{hatsuda} as 
\begin{equation}
\phi _{\alpha }\equiv \left\langle {\cal O} _{\alpha }\left
( 0,\vec{x}\right) \right\rangle =\frac{{\rm Tr}\left
[ {\cal O} _{\alpha }\left( 0,\vec{x}\right) e^{-\beta \left
( {\cal H}+{\cal H}_{ex}\right) }\right] }{{\rm Tr}\left[ e^{-\beta 
\left( {\cal H}+{\cal H}_{ex}\right) }
\right] }=\frac{1}{V}\int {\rm d}^{3}x\, \left\langle {\cal O} _{\alpha }
\left( 0,\vec{x}\right) \right\rangle \: , \label{eq1}
\end{equation}
where translational invariance is assumed and 
\({\cal H}_{ex} \) is given by
\begin{equation}
{\cal H}_{ex}=-\sum _{\alpha }\int
 {\rm d}^{3}x\, {\cal O} _{\alpha }\left( 0,
\vec{x}\right) {\cal F}_{\alpha }\: .\label{eq2}
\end{equation}

The (static) susceptibility \( \chi _{\alpha \beta } \) is defined as
\begin{eqnarray}
\chi _{\alpha \beta }(T) & = & \left. \frac{\partial \phi _{\alpha }}
{\partial {\cal F}_{\beta }}\right| _{{\cal F}=0}\nonumber \\
 & = & \beta \int {\rm d}^{3}x\, \langle {\cal O} _{\alpha }
( 0,\vec{x}) {\cal O} _{\beta }( 0,\vec{0}) \rangle \: , \label{eq3}
\end{eqnarray}
assuming no broken symmetry \( \left\langle {\cal O} _{\alpha }
\left( 0,\vec{x}\right) \right\rangle =\langle 
{\cal O} _{\beta }( 0,\vec{0}) \rangle =0 \). The quantity
$\langle {\cal O}_\alpha (0,{\vec x}){\cal O}_\beta(0,{\vec 0})\rangle $
is the two point correlation function with operators evaluated
at equal times.

\subsection{Chiral Susceptibility}

The chiral susceptibility measures the response of the chiral
condensate to the infinitesimal change of the current quark mass 
$m+\delta m$. Here, ${\cal F}_\alpha$ corresponds to 
the current quark mass and the operator ${\cal O}_\alpha$ to ${\bar q}q$. 
Then the {\it static} chiral
susceptibility can be obtained~\cite{hatsuda} from
\begin{equation}
\chi_c(T)= -\left.
\frac{\partial \langle {\bar q}q\rangle}{\partial m}\right |_{m=0}
=\beta\int {\rm d}^3x \ \left \langle {\bar q}(0,{\vec x})q(0,{\vec x}) 
{\bar q}(0,{\vec 0})q(0,{\vec 0}) \right \rangle 
=\beta \int {\rm d}^3x \; S(0, {\vec x})\ , \label{eq8}
\end{equation} 
where $S(0,{\vec x}) = \langle\cdots\cdots\rangle$ is the 
static correlator of the scalar channel. The chiral 
condensate is given as
\begin{equation}
{\langle {\bar q}q\rangle} =
\frac{{\rm{Tr}}\left [  {\bar q}q e^{-\beta {\cal H}}\right ]}
{{\rm{Tr}}\left [e^{-\beta {\cal H}}\right ]} \ =\frac{\partial\Omega}
{\partial m} \ .
\label{eq9}
\end{equation}
Here, $\cal H$ is the dynamical Hamiltonian of the system containing 
the quark mass, $\Omega=-(T/V)\log {\cal Z}$ is the thermodynamic potential, and
$\cal Z$ is the partition function of a quark-antiquark gas. 

Taking the Fourier transform of the static correlator 
$S(0,\vec{x})=\langle \cdots\cdots \rangle$,
it can be shown that
\begin{equation}
\chi_c(T) =\lim_{p\rightarrow 0}\> \beta \int_{-\infty}^{+\infty} 
\frac{{\rm d}\omega}{2\pi} S(\omega,{\vec p}) \ ,
\end{equation}
where $S(\omega,\vec p)$ is the Fourier transformed correlator.

Applying the fluctuation-dissipation theorem~\cite{hatsuda}, one  
finds the static limit of the {\it dynamic} chiral susceptibility
\begin{equation}
\tilde{\chi}_{c}\left( T\right) =\lim_{p\rightarrow 0} \> 
\beta \int _{-\infty }^{+\infty }\frac{{\rm d}\omega }
{2\pi }\, \frac{-2}{1-e^{-\beta \omega }}\, {\rm{Im}}\, 
\Pi \left( \omega ,{\vec p}\right) \, , \label{eq10}
\end{equation}
where $\Pi(\omega, {\vec p})$ is the scalar self-energy.
Using ${\rm{Im}}\Pi(-\omega, {\vec p})=-{\rm{Im}}\Pi(\omega, {\vec p})$, 
(\ref{eq10}) can also be written as
\begin{equation}
\tilde{\chi}_c(T)=-\lim_{p \rightarrow 0} \, \,
\frac{\beta}{\pi} \int_0^\infty {\rm d}\omega \ \coth 
\left( \frac{\beta  \omega}{2}\right ) \  {\rm{Im}} 
\Pi(\omega,{\vec p}) \ \ . \label{eq10a}
\end{equation}

\section{Perturbation Theory}

\subsection{Static Susceptibility}

The chiral condensate $\langle {\bar q}q\rangle$ at finite temperature
can be obtained from the tadpole diagram in Fig.1 as
\begin{eqnarray}
\langle {\bar q}q\rangle &=& - N_c\ N_f \ T \sum_{k_0} 
\int\frac{{\rm d}^3k}{(2\pi)^3} 
{\rm{Tr}} \Big [ S(K)\Big ] \, , \label{eq11}
\end{eqnarray}
where, $N_f$ and $N_c$ are, respectively, the numbers of quark flavors 
and colors.
$S(K)$ is the free quark propagator given as 
\begin{equation}
S(K)=\frac {1}{\slash \hspace{-0.25cm}K -m} \ \ .
\end{equation}
Summing over the Matsubara frequencies $k_0$, we obtain
\begin{eqnarray}
 \langle {\bar q}q\rangle
&=& - 2N_c \ N_f \ \int \frac{{\rm d}^3k}{(2\pi)^3} \frac{m}{E_k} \Big 
[ 1 - 2n_F(E_k) \Big ] \ , \label{eq11b}
\end{eqnarray}
where $n_F(E_k)$ is the Fermi distribution function and $E_k=\sqrt{k^2+m^2}$. 

\vspace{0.5cm}
\begin{center}
\begin{picture}(300,100)(0,0)
\ArrowArcn(150,50)(40,270,90)
\ArrowArcn(150,50)(40,90,270)
\DashArrowLine(50,10)(150,10){5}
\DashArrowLine(150,10)(230,10){5}
\Text(150,103)[]{$K$}
\end{picture}
\end{center}
\centerline{Fig.1: Tadpole diagram determining the free static chiral
susceptibility.}
\vspace{0.3in}

Alternatively, the chiral condensate can be derived from the lowest order
thermodynamic potential of the quark-antiquark system~\cite{kapus}
\begin{equation}
\Omega = - 2N_fN_c T \int \frac{{\rm d}^3k}{(2\pi)^3}\left \{ \beta E_k +
\log \left [1 + e^{-\beta(E_k-\mu_q)}   \right ]
+\log \left [1 + e^{-\beta(E_k+\mu_q)}   \right ] \right \} \ . \label{eq12}
\end{equation}
Using (\ref{eq9}) the condensate (\ref{eq11b}) is reproduced.

Now, the chiral susceptibility in the free case 
follows from (\ref{eq8}) and 
(\ref{eq11b}) as
\begin{eqnarray}
\chi_c^f(T)= -\left. \frac{\partial \langle {\bar q}q\rangle}{\partial m}
\right|_{m=0}  
&=&2N_f N_c \int \frac{{\rm d}^3k}{(2\pi)^3} \left. \left \{ \frac{k^2}{E_k^3} 
\Big [1-2n_F(E_k)\Big ]
+2\beta \frac {m^2}{E_k^2}n_F(E_k)\Big[1-n_F(E_k)\Big]
\right \}\right |_{m=0}
 \nonumber \\
&=&2N_f N_c \int \frac{{\rm d}^3k}{(2\pi)^3}  \frac{1}{k} 
\Big[1-2n_F(k)\Big]  \, . 
 \label{eq14} 
\end{eqnarray}

Note that, in contrast to the quark number susceptibility~\cite{mun} where the
zero temperature contribution vanishes due to the transversality properties of
the vector channel \cite{mclerran}, the static
chiral susceptibility contains a quadratic ultraviolet divergence coming from the
zero temperature contribution. Using dimensional regularization the temperature 
independent term in (\ref{eq14}) disappears, 
as there is no scale associated with it, 
leading to

\begin{equation}
\chi_c^f(T)=-\frac{N_fN_c}{6}\> T^2.
\label{eq14a}
\end{equation}

The next order contribution follows from the two-loop thermodynamic potential given in 
Ref.~\cite{kapus}. However, the second derivative of this expression with respect to $m$ 
diverges at $m=0$ (see also Ref.~\cite{muk}). 
The reason for this divergence can be seen in the following way: the second
derivative with respect to the quark mass $m$ of the two-loop thermodynamic potential, which is 
a quark loop with an internal gluon line, introduces two additional quark propagators. Hence
the originating diagram looks like the polarization tensor containing an internal quark line.
This diagram is known to be logarithmically infrared divergent in the case of a vanishing bare 
quark mass (see e.g. \cite{KLS}).
Hence, the static chiral susceptibility cannot be calculated consistently 
in usual perturbation theory beyond leading order but requires the HTL resummation.  

\subsection{Dynamic Susceptibility in the Static Limit}

Next, we calculate the static limit of the dynamical chiral susceptibility
from the lowest order self-energy diagram shown in Fig.2:
 \begin{equation}
\Pi \left( P\right) =N_{f}N_{c}T\sum _{k_{0}}\int \frac{{\rm d}^{3}k}{(2\pi )
^{3}}{\rm {Tr}} \Big[ S\left( K\right) S\left( Q\right)
\Big] \, . \label{eq15}
\end{equation}
Summing over $k_0$, we obtain
\begin{eqnarray}
{\rm{Im}} \Pi(\omega, {\vec p}) &=& -N_f N_c \pi \int \frac{{\rm d}^3k}
{(2\pi)^3}
\left \{ \left (1-\frac{{\vec k}\cdot {\vec q}+ m^2}{E_kE_q} \right )
\Big [1-n_F(E_k)-n_F(E_q) \Big ]
\Big [ \delta(\omega -E_k-E_q) - \delta (\omega+E_k+E_q) \Big ] \right.
\nonumber \\
&+&  
\left.  \left (1+\frac{{\vec k}\cdot {\vec q}+ m^2}{E_kE_q} \right )
\Big [n_F(E_k)-n_F(E_q) \Big ]
\Big [ \delta(\omega -E_k+E_q) - \delta (\omega+E_k-E_q) \Big ] \right \}
\label{eq16} 
\end{eqnarray}

\vspace*{0.5cm}
 \begin{center}
\begin{picture}(300,100)(0,0)
\ArrowArc(150,50)(40,180,0)  
\ArrowArcn(150,50)(40,180,0) 
\DashArrowLine(50,50)(110,50){5}
\DashArrowLine(190,50)(250,50){5}
\Text(75,40)[]{$P$}
\Text(225,40)[]{$P$}
\Text(150,105)[]{$K$}
\Text(150,-5)[]{$Q=P-K$}
\end{picture}
\end{center} 
\vspace{0.2in}
\centerline{Fig.2: Self-energy diagram determining the free dynamic chiral
susceptibility.}
\vspace{0.3in}

Combining (\ref{eq10}) and (\ref{eq16}) yields 
\begin{eqnarray}
\tilde{\chi}_c^f(T)&=&2N_f N_c \int \frac{{\rm d}^3k}{(2\pi)^3}\left.
 \left \{ \beta \frac{k^2}{E_k^2} 
\Big [1-2n_F(E_k)+2n_F^2(E_k)\Big ] 
+2\beta \frac {m^2}{E_k^2}n_F(E_k)\Big [1-n_F(E_k)\Big ]\right \}\right |_{m=0}
 \nonumber \\
&=&2N_f N_c \beta \int \frac{{\rm d}^3k}{(2\pi)^3}   
\Big [1-2n_F(k)+2n_F^2(k)\Big ] \ . 
\label{eq17} 
\end{eqnarray}
Obviously (\ref{eq14}) and (\ref{eq17}) do not agree. For, in 
general, the static susceptibility, given in (\ref{eq14}) 
as the second derivative of the thermodynamical potential, does not
coincide with the static limit 
of the dynamical susceptibility in (\ref{eq17}) which satisfy the 
fluctuation-dissipation theorem.
There is an exception: if the static susceptibility 
were associated with a conserved quantity such as the (quark) number density,
the conservation law would ensure the equality of the two different
quantities~\cite{hatsuda,fluc}. Thus, as shown in Refs.~\cite{hatsuda,mun},
the quark number susceptibility can be derived in two different ways as the
imaginary part of the self-energy is proportional to $\delta(\omega)$ implying a 
conserved number density.
However, the static chiral susceptibility is given by the derivative
of the scalar/chiral density in (\ref{eq14}) which obeys no conservation 
law, so it does not coincide with the static limit of the dynamic 
susceptibility in (\ref{eq17}) as represented by the fluctuation-dissipation 
theorem since the imaginary part of the self energy in the static limit
is not proportional to $\delta(\omega)$. 

Note that, in contrast to the static chiral susceptibility, there is no zero temperature
contribution as there is an overall factor $\beta$. The term containing no 
distribution function in (\ref{eq17}) shows a cubic ultraviolet divergence.

\section {HTL Approach}

Here, we want to calculate the chiral susceptibility beyond the free
quark approximation by taking into account the in-medium properties of quarks in a
QGP. In the weak coupling limit ($g\ll 1$), a consistent method is to use the
HTL-resummed quark propagator and HTL quark-meson vertex\footnote{In the case of a scalar meson
there is no HTL contribution to the quark-meson vertex \cite{karsch}.} if the quark
momentum is soft ($\sim gT$). However, in the following we will use the
HTL-resummed quark propagator also for hard momenta, in order to
consider, at least to some extent, non-perturbative features of the QGP such as
effective quark masses and Landau damping. This strategy is in the same spirit as 
in the case of the quark number susceptibility \cite{mun} and the meson correlators
\cite{karsch}, although this approach might not be consistent in the weak coupling 
limit \cite{blaizot}. However, since we do not want to restrict to the weak coupling limit, 
but consider temperatures close to $T_c$, we do not aim at a complete leading 
order calculation, using here
the HTL Green functions only for including medium effects in a gauge invariant
way.

Another, maybe more convincing, reason for studying the chiral susceptibility in 
the HTL approximation is that this approach has also been
used for the free energy of the QGP, where different implementations of the HTL method
and their validity have been discussed. The chiral susceptibility may serve as another
test for these questions.

In the rest frame of the medium the most general expression for the fermion 
self energy in the chiral limit of vanishing current mass has the 
form~\cite{weld}
\begin{equation}
\Sigma(K)=a \slash {\hspace{-0.25cm}}K -b\gamma_0 , \label{2.5} 
\end{equation}
where the scalar quantities $a$ and $b$ are functions of energy $k_0$ and 
the magnitude $k$ of the three momentum.
The HTL-resummed quark propagator, $S^\star(K)=1/[\slash\hspace{-0.25cm}K-
\Sigma(K)]$, can be expressed
using the helicity representation in the massless case as \cite{bpy}
 \begin{equation}
\label{2.6}
S^\star\left( k_{0},k\right) =\frac{\gamma ^{0}-\widehat{k}.
\vec {\gamma }}{2D_{+}\left( K\right) }+\frac{\gamma ^{0}
+\widehat{k}.\vec{\gamma }}{2D_{-}\left( K\right) }
\end{equation}
with
\begin{eqnarray}
D_{\pm }\left( k_0,k\right) &=& \left (-k_0\pm k\right )  
\left (1+a\right ) - b \nonumber \\
&=& -k_0 {\pm} k+\frac{m^{2}_{q}}
{k}\left[ Q_{0}\left( \frac{k_0}{k}\right) \mp Q_{1}\left( \frac{k_0}{k}
\right) \right] \, , \label{2.7}
\end{eqnarray}
where the thermal quark mass is given by
$m_q=g(T) T/{\sqrt 6}$, and $Q_n(y)$ is the Legendre
function of second kind. 
The zeros $\omega_\pm(k)$ of $D_\pm(K)$ describe the two dispersion
relations of collective quark modes in a thermal medium~\cite{bpy}.
Furthermore, the
HTL-resummed quark propagator acquires a cut contribution below the
light cone ($k_0^2 < k^2$) as the HTL quark self-energy, contained in
$D_\pm(K)$, has a non-vanishing
imaginary part, which can be related to Landau damping for spacelike quark
momenta resulting from interactions of the quarks with gluons of the
thermal medium. For the coupling constant entering $m_q(T)$, 
$\alpha_s=g^2/(4\pi)$, we choose 
the two-loop result of the temperature dependent running 
coupling\footnote{In this way, we have combined an infrared
(HTL) and an ultraviolet (renormalization group) resummation scheme
phenomenologically. A systematic combination of these techniques is not
available yet.} \cite{andersen00} 
\begin{equation}
\alpha_s(\mu_4) = \frac{4\pi}{\beta_0{\bar L}} \left [ 1 -
\frac{2\beta_1}{\beta_0^2} \frac{\log {\bar L}}{{\bar L}}\right ] \, ,
\label{coupling}
\end{equation}
where $\beta_0=(11- \frac{2}{3}N_f)$, $\beta_1=(51-\frac{19}{3}N_f)$,
${\bar L}=\log\left (\mu_4^2/\Lambda_{\overline{\rm{MS}}}^2\right)$ and 
$\mu_4 \sim T$ is the renormalization scale for the running coupling constant. 
It should also be noted that the
HTL-resummed propagator is chirally symmetric in spite of the appearance
of an effective quark mass~\cite{weld}.

The HTL-spectral function of $1/D_\pm(K)$, $\rho_\pm = -{\rm Im}(1/D_\pm)/\pi$,
can be written as \cite{bpy}
\begin{equation}
\rho_\pm(k_0,k) = \frac{k_0^2-k^2}{2m_q^2} \biggl [ \delta(k_0-\omega_\pm)
+\delta(k_0+\omega_\mp) \biggr ]
+\beta_\pm(k_0,k) \Theta(k^2-k_0^2) \label{2.10}
\end{equation}
with
\begin{equation}
\beta_\pm(k_0,k) = -\frac{m_q^2}{2} \frac{\pm k_0 -k}
{\biggl [  k (-k_0 \pm k) +m_q^2\left (\pm 1- \frac{\pm k_0 -k}{2k} \log
\frac{k+k_0}{k-k_0} \right )\biggr ]^2 + \biggl[ \frac{\pi}{2} m_q^2
\frac{\pm k_0 -k}{k} \biggr ]^2} \, \, \, ,  \label{2.11}
\end{equation}
where the first term describing the quasiparticle dispersion
in (\ref{2.10}) is due to the poles of the propagator
and the second term containing Landau damping
corresponds to the cut contribution for spacelike
quark momenta.

\subsection{Static Susceptibility}

Now, we compute the chiral condensate $\langle q\bar q\rangle$ from the
tadpole diagram in Fig.1, where we use the effective HTL quark propagator,
yielding 
\begin{eqnarray}
\langle {\bar q}q\rangle &=& - N_c\ N_f \ T \sum_{k_0} 
\int\frac{{\rm d}^3k}{(2\pi)^3} 
{\rm{Tr}} \Big [ S^\star (K)\Big ]. \label{2.18}
\end{eqnarray}
A similar calculation has been done in the case of the quark contribution
to the free energy by Andersen et al. \cite{andersen00}.

According to (\ref{eq8}), but in contrast to the HTL approximation discussed 
above, we have to
take into account a non-vanishing current quark mass. 
In the weak coupling limit (HTL approximation) there are two possibilities:
1) $m \geq T$,  i.e., $m \gg \Sigma_{\rm{HTL}} \sim gT$ and 2) $m\leq gT$. In 
the first case, 
$\Sigma_{\rm{HTL}}$ can 
be neglected in the effective quark propagator compared to the hard scale 
$m$, implying 
$S^\star (K) = S(K)$, which leads to the free susceptibility. In the 
second case, 
$m$ can be neglected in the calculation of the
HTL quark self energy, requiring hard loop momenta, i.e.,  
$\Sigma \simeq \Sigma_{\rm{HTL}}(m=0)$. 
Then, the effective HTL quark propagator becomes 
\begin{eqnarray}
S^\star(K)= \frac{1}{\slash {\hspace{-0.25cm}}K(1+a)+b\gamma_0 -m}
       =\frac{\slash {\hspace{-0.25cm}}K(1+a)+b\gamma_0 +m}{D-m^2}
\label{2.20} 
\end{eqnarray}
with
\begin{equation}
D(K) = K^2 (1+a)^2+ 2k_0b(1+a)+b^2=D_+(K)D_-(K), \label{2.21}
\end{equation}
where $D_\pm(K)$ is given by the massless HTL-approximation in (\ref{2.7}).
Note that the current quark mass, $m\leq gT$, is of the same order as the HTL
contributions $a$ and $b$ and cannot be neglected therefore.

Summing over $k_0$, we obtain
\begin{equation}
\langle q \bar q \rangle = 4mN_fN_c \int \frac {{\rm d}^3k}{(2\pi)^3}
\int_{-\infty}^\infty {\rm d}x 
\rho(x, k, m) \left [ 1-n_F(x) \right ] \ ,
\label{2.22}
\end{equation}
where the spectral function corresponding to the massive HTL propagator is
given by
\begin{equation}
\rho(x,k,m) = -\frac{1}{\pi} {\rm{Im}} \frac{1}{D-m^2} \ . \label{2.23}
\end{equation}

Following (\ref{eq8}), the static chiral susceptibility in 
HTL-approximation is found as
\begin{eqnarray}
\chi_c^h(T) &=& -\frac {\partial \langle q\bar q\rangle}{\partial m}\nonumber \\
&=& -4N_fN_c \int \frac {{\rm d}^3k}{(2\pi)^3} \biggl\{
\frac {\omega^2_+(k)-k^2 }{2m_q^2} 
{\rm Re}D_-^{-1}(\omega_+,k)\left [1-2n_F(\omega_+)\right ]
+ \frac {\omega^2_-(k)-k^2 }{2m_q^2} 
{\rm Re}D_+^{-1}(\omega_-,k)\left [1-2n_F(\omega_-)\right] \nonumber \\
&+& \int_{0}^k {\rm d}x \> [\beta_+(x,k){\rm Re}D_-^{-1}(x,k)+
\beta_-(x,k){\rm Re}D_+^{-1}(x,k)]
\left [1-2n_F(x)\right ] \biggr\} \ , \label{2.24}
\end{eqnarray}
where
\begin{equation}
{\rm Re}D_\pm^{-1}(k_0,k) = \frac{k^2(-k_0 \pm k)+{m_q^2k}\left (\pm 1- \frac{\pm k_0 -k}{2k} \log
\left |\frac{k+k_0}{k-k_0} \right |\right )}
{\biggl [  k (-k_0 \pm k) +m_q^2\left (\pm 1- \frac{\pm k_0 -k}{2k} \log
\left |\frac{k+k_0}{k-k_0} \right |\right )\biggr ]^2 + \biggl[ \frac{\pi}{2} m_q^2
\frac{\pm k_0 -k}{k} \theta (k^2-k_0^2)\biggr ]^2} \, \, \, .  \label{2.24a}
\end{equation}
In the infinite temperature limit, the first term in (\ref{2.24}) reduces to the free
static susceptibility given in (\ref{eq14}) whereas the second and third terms 
vanish.
The integrals without the distribution functions in the first and third term
in (\ref{2.24}) are ultraviolet divergent (quadratically and logarithmically) because of the 
asymptotic behavior of the
quasiparticle dispersion relation $\omega_+(k)$, ${\rm Re}D_\pm^{-1}$, and $\beta_\pm$. 
However, the second term is convergent since the plasmino dispersion relation,
$\omega_-(k)$, approaches $k$ exponentially at large $k$. The quadratic divergence is temperature 
independent and can be removed by subtracting the zero temperature part, coinciding with dimensional 
regularization, or by subtracting the free chiral susceptibility (\ref{eq14}) as discussed below. 
The remaining logarithmic divergence turns out to be temperature dependent as in the case of the 
gluonic part of the free energy \cite{andersen00}. This might be an artifact of the approximation 
used here and might vanish if one goes beyond the one-loop HTL approximation.
Here we will use dimensional regularization to remove this divergence following Ref.~\cite{andersen00}.

First we subtract the free  
susceptibility given in (\ref{eq14}) from the HTL one in (\ref{2.24}): 

\begin{eqnarray}
{ \chi}_c^{{h}} - { \chi}_c^{ f} &=&
\frac{2N_fN_c}{\pi^2} \int_{0}^\infty {\rm d}k \, k \, n_F(k) 
 +\frac{4N_fN_c}{\pi^2} \int_{0}^\infty {\rm d}k \, k^2 \, \frac{(\omega_+^2-k^2)}{2
m_q^2} \,
{\rm {Re}} D_-^{-1} (\omega_+,k) \, n_F(\omega_+) \nonumber \\
&& -\frac{2N_fN_c}{\pi^2} \int_{0}^\infty {\rm d}k \, k^2 \, \frac{(\omega_-^2-k^2)}{2
m_q^2} \,
{\rm {Re}} D_+^{-1} (\omega_-,k) \,\left (1-2n_F(\omega_-)\right ) \nonumber \\
&&- \frac{N_fN_c}{\pi^2} \int_{0}^\infty {\rm d}k \, \left [ k \, +\, 2k^2 
\frac{(\omega_+^2-k^2)}{2m_q^2}
{\rm {Re}} D_-^{-1} (\omega_+,k) \right ] \nonumber \\
&&+ \frac{4N_fN_c}{\pi^2} \int_{0}^\infty {\rm d}k \, k^2 \, \int_{0}^k dx \, 
\left [ \beta_+(x,k) \, {\rm {Re}} D_-^{-1} (x,k) \, n_F(x) 
+    \, 
\beta_-(x,k) \, {\rm {Re}} D_+^{-1} (x,k) \, n_F(x) \right ] \nonumber \\
&&- \frac{2N_fN_c}{\pi^2} \int_{0}^\infty {\rm d}k \, k^2 \, \int_{0}^k dx \,
\left [ \, \beta_+(x,k) \,
 {\rm {Re}} D_-^{-1} (x,k) \, + \,  \beta_-(x,k) \,
 {\rm {Re}} D_+^{-1} (x,k)\right ]. \label{fsub}
 \end{eqnarray}
The integrals in 4$^{\rm {th}}$ and 6$^{\rm {th}}$ terms are, respectively, 
logarithmically ultraviolet divergent. 
Using a momentum cutoff $k < \Lambda$ and energy cutoff $x < \Lambda$,
the 4$^{\rm {th}}$ and 6$^{\rm {th}}$ terms can be seen to be proportional to $m_q^2 \log \Lambda$. 
This is in contrast to the quark contribution to the
free energy but similar to the case of the gluonic part of the free energy where these 
terms do 
not cancel~\cite{andersen00}. Dimensional regularization will then replace the remaining
logarithmic divergences by poles in $d-3$, which finally will require a counter term
for cancellation~\cite{andersen00}. 

As pointed out in Ref.~\cite{Blaizot01} it is important to use also $d$-dimensional
expressions for the HTL propagator. Here we will follow Ref.\cite{Andersen02,Andersen02a}
and expand the chiral susceptibility up to order $m_q^4$. 
The angular average leading to the Legendre functions in (\ref{2.7}) 
can be generalised to $d=3-2\epsilon$ spatial dimensions in 
Minkowski space as
\begin{eqnarray}
{\cal T}_d (K) &=& \frac{w(\epsilon)}{2} \int_{-1}^{1} dc \ \
(1-c^2)^{-\epsilon} \frac{k_0}{k_0-kc}
, \label{1} 
\end{eqnarray}
where the weight function $w(\epsilon)$ is
\begin{equation}
w(\epsilon) = \frac{\Gamma(\frac{3}{2}-\epsilon)}{\Gamma(\frac{3}{2})
\Gamma(1-\epsilon)} \ . \label{1a}
\end{equation}  
Eq.~(\ref{1}) can also be written as~\cite{Andersen02}
\begin{eqnarray}
{\cal T}_d (K) &=& w(\epsilon) \int_{0}^{1} dc \ \
(1-c^2)^{-\epsilon} \frac{k_0^2}{(k_0^2-k^2c^2)}
=\left \langle \frac{k_0^2}{k_0^2-k^2c^2} \right \rangle_c.
\label{1b} 
\end{eqnarray}

Now, it is convenient to define the functions $A_0(K)$ and $A_S(K)$ as
\begin{eqnarray} 
A_0 &=& k_0 -\frac{m_q^2}{k_0} {\cal T}_d(K) \ , \nonumber \\
A_S &=& k +\frac{m_q^2}{k} \left (1- {\cal T}_d(K) \right ) , \label{3}
\end{eqnarray} 
which are related to the HTL-resummed quark propagator (\ref{2.7}) by 
\begin{equation}
D_{\pm }\left( k_0,k\right) = - \left (A_0\mp A_S \right ). 
\label{5}
\end{equation}

Using (\ref{eq8}), (\ref{2.18}), and (\ref{2.20}) the chiral susceptibility in one-loop HTL order reads
\begin{equation}
\chi_c^h(T) = -4 N_f N_c \SI \frac{1}{D(K)} , \label{6}
\end{equation}
where \cite{Andersen02a}
\begin{equation}
\SI = \left (\frac{e^\gamma \mu^2}{4\pi}\right )^\epsilon \> (-iT)\> \sum_{k_0}\> \int \frac{d^{3-2\epsilon}p}
{(2 \pi)^{3-2\epsilon}}
\end{equation}
with the arbitrary renormalization scale $\mu $ of the (${\overline {\rm {MS}}}$) renormalization scheme  
and Euler's constant $\gamma$. 

Expanding $\chi_c^h(T)$ in a power series of $m_q^2/T^2$, and 
keeping terms up to
second order, we obtain
\begin{eqnarray}
\chi_c^h(T) &=&-4N_fN_c \SI \ \left [ \frac{1}{K^2} + 2 m_q^2\frac{1}{K^4}
+m_q^4\left ( \frac{4}{K^6} +\frac{1}{k^2K^4}-\frac{2}{k^2K^4}{\cal T}_d(K)
+\frac{1}{k^2k_0^2 K^2}{\cal T}_d^2(K)\right )\right ] . \label{8}
\end{eqnarray} 
Using the sum-integrals evaluated in Appendix A, we end up with
\begin{eqnarray}
\chi_c^h(T) &=&-\frac{N_fN_c}{6}T^2 \left[ 1 +\frac{3}{\pi^2}
\left (\frac{\mu}{4 \pi T}\right )^{2\epsilon} \left(\frac{1}{\epsilon}
+2\gamma+4\log 2 \right) \frac{m_q^2}{T^2} +\frac{21 \zeta(3)}{4 \pi^4}
\left(5+\frac{8}{7}\log 2 -\frac{\pi^2}{6} \right) \frac{m_q^4}{T^4}\right ]
 \label{9}
\end{eqnarray}
The chiral susceptibility is logarithmic divergent at order $m_q^2$, but 
finite at order $m_q^4$. We choose a counter term within the 
minimal subtraction renormalization scheme just to cancel the pole in 
$1/\epsilon$:
\begin{equation}
\Delta \chi_c^{\rm{counter}}=\frac{N_f N_c}{2\pi^2 \epsilon}m_q^2 . \label{10}
\end{equation}
The complete expression for chiral susceptibility is then given by
\begin{eqnarray}
\chi_c^h(T) &=&\chi_c^f(T) \left[ 1+\frac{6}{\pi^2}
\left (\log\frac{\mu}{4 \pi T}+\gamma +2\log 2 \right) \frac{m_q^2}{T^2}
+\frac{21}{4}\frac{\zeta(3)}{\pi^4}
\left(5+\frac{8}{7}\log 2 -\frac{\pi^2}{6} \right)
\frac{m_q^4}{T^4}
\right ] . \label{11}
\end{eqnarray}

\centerline{\psfig{figure=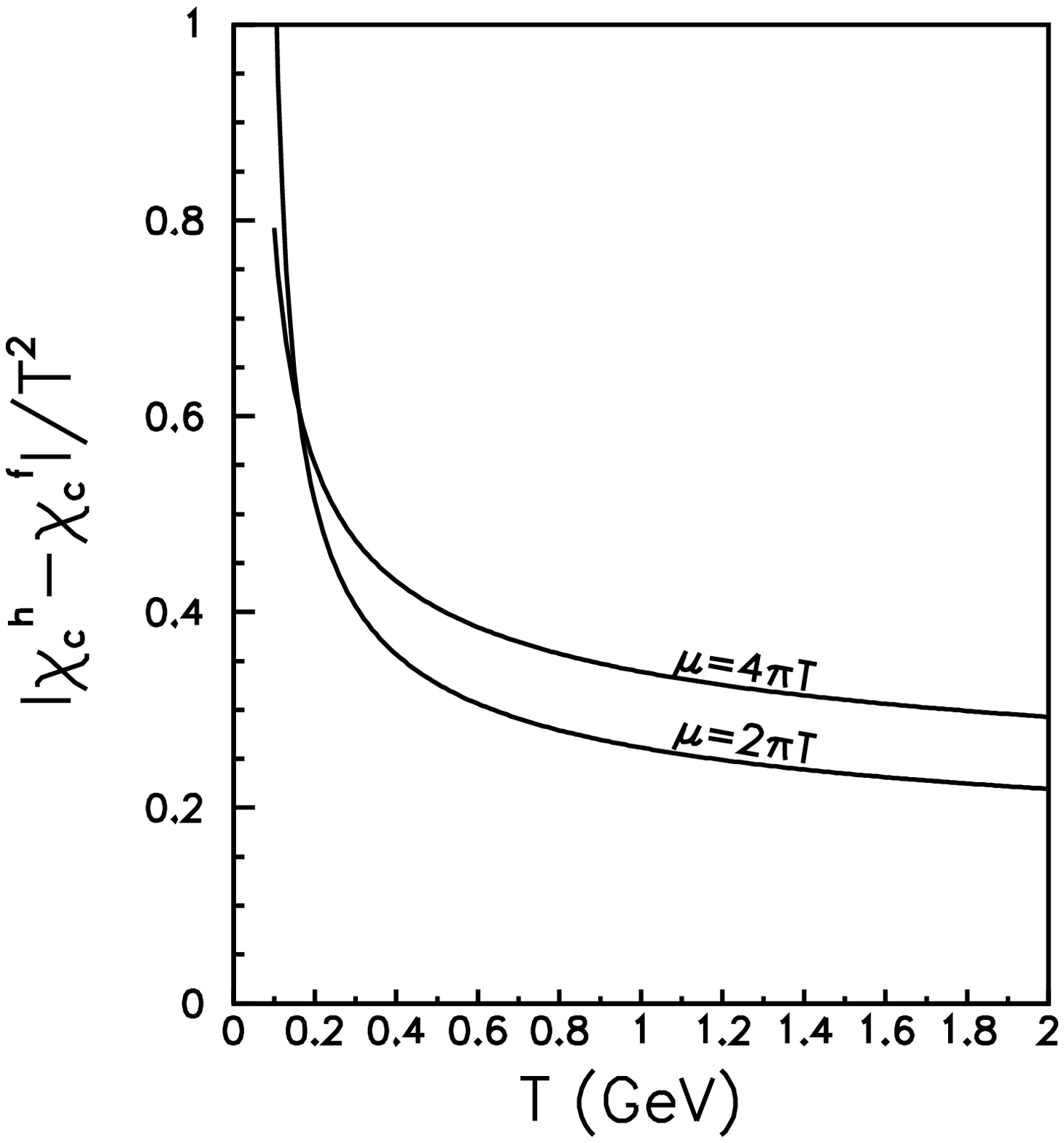,height=10cm}\hspace*{0.5cm}
\psfig{figure=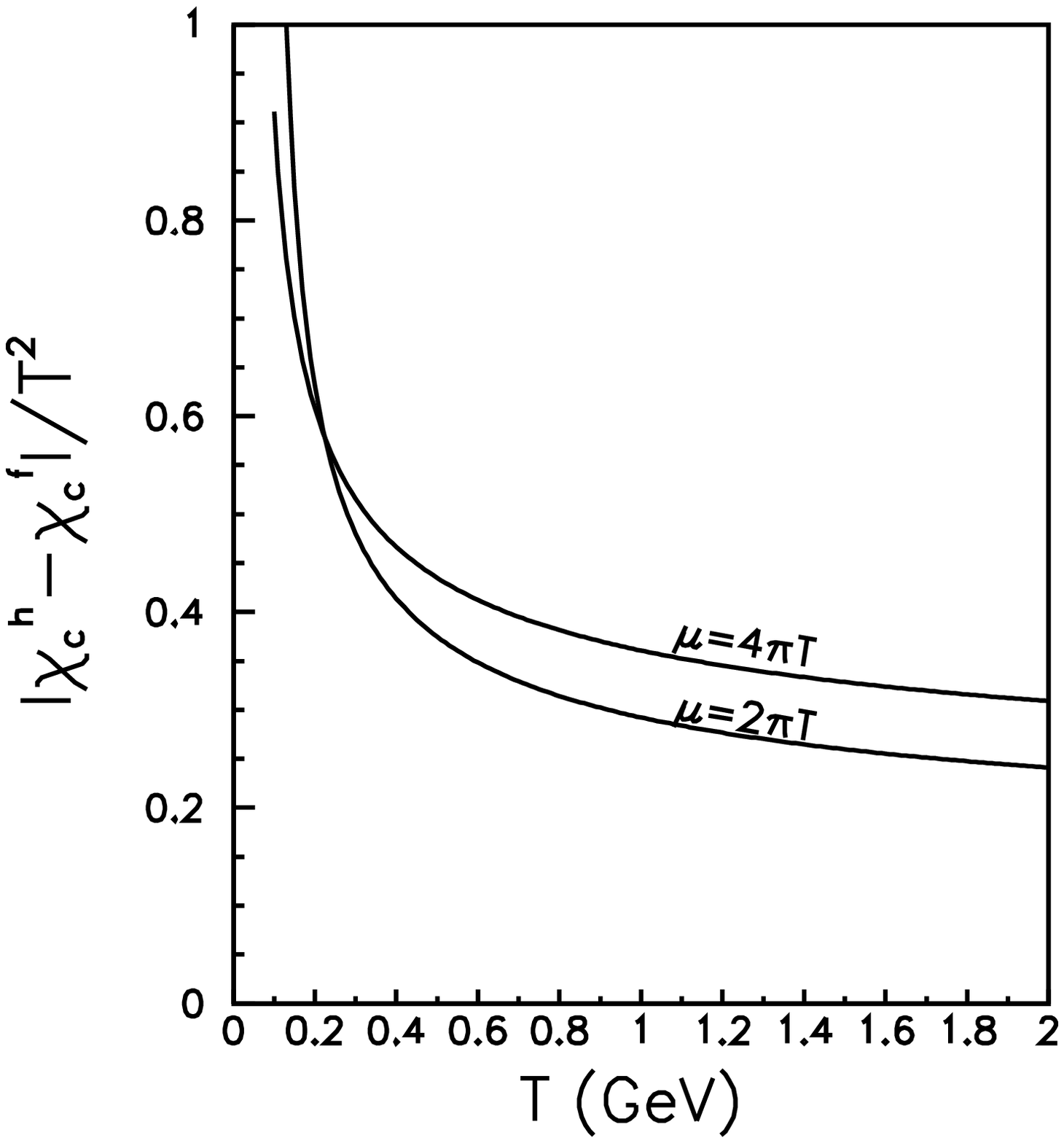,height=10cm}}

\vspace*{-1.5cm}

\noindent {Fig 3: The HTL static susceptibility (free susceptibility subtracted)  
as a function of temperature for $\Lambda_{\overline{\rm{MS}}}=300$ MeV,
and $N_f=2$ with the choices of the renormalization scale 
$\mu=2\pi T$ and $4 \pi T$ for up to order $m_q^2$ (left) and
up to order $m_q^4$ (right).}

\vspace*{0.5cm}

Note that the $m_q^2$-term is the first finite $\alpha_s$-correction to the free chiral 
susceptibility, since the two-loop contribution in usual perturbative QCD diverges for $m=0$.

The quantity $|\chi_c^h(T)-\chi_c^f(T)|/T^2$ as a function of $T$
is shown in Fig.3 for two different choices of $\mu$, 
where we have also chosen $\mu_4=\mu$ in $\alpha_s$ entering $m_q$. Obviously the $m_q^4$ 
contribution is only a small correction. Whereas in the $m_q^2$
contribution only the effective quark mass (pole contribution) 
enters, there is a Landau damping contribution coming from
${\cal T}_d$ in the $m_q^4$ term (see (\ref{8})). 
To understand the behavior of the susceptibility in Fig.3 we ascertain that
the quantity $|\chi_c^h-\chi_c^f|/T^2$, plotted in Fig.3, depends on the temperature only via
the temperature dependent running coupling constant, i.e. $|\chi_c^h-\chi_c^f|/T^2=F[g(T)]$.  
This can be seen by scaling the variables $k$ and $x$ in the integrals of (\ref{2.24}) by $m_q$.
It holds also after dimensional regularization, if a renormalization scale proportional to the 
temperature is adopted in the latter case (see e.g. (\ref{11})). 
This dependence has also been verified
numerically, where a constant, temperature independent chiral susceptibility 
has been observed, if a constant, temperature independent coupling has been employed.
In other words, the decrease of the susceptibility with temperature in Fig.3 
is entirely caused by the temperature dependence of the coupling constant.
Due to asymptotic freedom, $g(T\rightarrow \infty)\rightarrow 0$, the HTL susceptibility 
reduces to the free one in the infinite temperature limit, i.e. $F[g=0]=0$.

In lattice simulations a peak around the critical temperature associated with chiral symmetry restoration
has been observed. In contrast to lattice QCD the HTL approximation, although taking into
account medium effects in the plasma due to interactions, does not contain any physics related to
the chiral restoration, which is a truly non-perturbative effect. 
In Fig.3 we observe a strong increase of the magnitude of the HTL chiral
susceptibility towards low temperatures, similar as in lattice simulations above $T_c$.
However, this increase is caused only by the fact that for our choice of the renormalization
constant $\mu$ the HTL result is above the free one, $|\chi_c^h|>|\chi_c^f|$, 
and that the temperature dependent coupling increases at smaller temperatures. 
If we choose a small enough value for $\mu$, $|\chi_c^h|<|\chi_c^f|$ holds according to (\ref{11}), and 
the HTL susceptibility $|\chi_c^h|$ approaches the free susceptibility $|\chi_c^f|$ 
from below at large temperatures. This indicates the strong sensitivity to the choice
of the regularization procedure and the renormalization constant similar as in the case of the free
energy \cite{Andersen02,Andersen02a}.
Hence we understand the difference between the lattice and the HTL approach qualitatively
but the different ultraviolet behavior of the HTL 
and the lattice result, which 
does not contain a temperature dependent logarithmic divergence \cite{private}, 
renders a direct, quantitative comparison of the both approaches difficult.

\subsection{Dynamic Susceptibility in Static Limit}

Here, we need to calculate the imaginary part of the self energy diagram
given in Fig.2, where both bare quark propagators are replaced by
effective HTL quark propagators.
It is sufficient to restrict to the massless case, i.e., (\ref{2.6}). 

Apart from constant factors the dynamic chiral susceptibility agrees with the
scalar meson correlator at $\tau = \beta$ [compare (2.3) of Ref.\cite{karsch}
with (\ref{eq10a})]. Hence, we will not repeat the derivation here but just
present the final results.

As in the case of the meson correlator \cite{karsch}, the dynamic susceptibility
can be decomposed into pole-pole, pole-cut and cut-cut contributions,
\begin{equation}
\tilde{\chi}_c(T) = \tilde{\chi}^{\rm{pp}}_c(T)
+\tilde{\chi}^{\rm{pc}}_c(T)+\tilde{\chi}^{\rm{cc}}_c(T)  . \label{2.14b}
\end{equation}
The pole-pole contribution  reads
\begin{eqnarray}
\tilde{\chi}^{\rm{pp}}_c(T) &=&
{2N_cN_f\beta}  \int \frac{{\rm d}^3k}{(2\pi)^3}
\biggl\{\frac{[ \omega_+^2(k) -k^2]^2}{4m_q^4}
 \Big [ 1 -2n_F(\omega_+) +2n_F^2(\omega_+)\Big ] 
\nonumber \\
&& + \frac{[ \omega_-^2(k) -k^2]^2}{4m_q^4}
 \Big [ 1 -2n_F(\omega_-) +2n_F^2(\omega_-)\Big ] 
\nonumber \\
&& + \frac{[ \omega_+^2(k) -k^2][ \omega_-^2(k) -k^2]}
{2m_q^4} \Big [n_F(\omega_+)[1-n_F(\omega_-)]+ 
n_F(\omega_-)[1-n_F(\omega_+] \Big ]
 \biggr\} \; . \label{2.15} 
\end{eqnarray}
In the infinite temperature limit the plasmino mode decouples from the medium,
causing the second and third terms to vanish, whereas the first term reduces to
the free susceptibility given in (\ref{eq17}).
The pole-cut contribution is given as 
\begin{eqnarray}
\tilde{\chi}_c^{\rm{pc}}(T) &=&
\frac{2N_fN_c\beta} {m_q^{2}} \int_0^\infty d\omega 
\coth \left(\frac{\beta \omega}{2}\right ) \int \frac{{\rm d}^3k}{(2\pi)^3}
\nonumber \\
&~& \bigl\{\Theta [k^2-(\omega-\omega_+)^2 ] n_F(\omega-\omega_+) 
n_F(\omega_+) \beta_+(\omega-\omega_+,k) (\omega_+^2-k^2) \nonumber\\
&~&+ \Theta[k^2-(\omega-\omega_-)^2]n_F(\omega-\omega_-)n_F(\omega_-) 
\beta_-(\omega-\omega_-,k) (\omega_-^2-k^2) \nonumber\\
&~&+\Theta [k^2-(\omega+\omega_-)^2] n_F(\omega+\omega_-) [1-n_F(\omega_-)] 
\beta_+(\omega+\omega_-,k) (\omega_-^2-k^2) \nonumber\\
&~&+\Theta [k^2-(\omega+\omega_+)^2] n_F(\omega+\omega_+) [1-n_F(\omega_+)] 
\beta_-(\omega+\omega_+,k) (\omega_+^2-k^2) \bigr\}\; , \label{2.16}
\end{eqnarray}
and the cut-cut contribution as
\begin{eqnarray}
\tilde{\chi}_c^{\rm{cc}} (T) &=& {2N_fN_c\beta} 
 \int_0^\infty d\omega \coth \left(\frac{\beta \omega}{2}\right ) 
\int \frac{{\rm d}^3k}{(2\pi)^3}
\int_{-k}^{k} {\rm d}x 
\; n_F(x) n_F(\omega-x) \Theta(k^2-(x-\omega)^2) 
\nonumber \\
&~&\times \biggl[ \beta_+(x,k) \beta_+(\omega-x,k) + \beta_-(x,k) 
\beta_-(\omega-x,k) \biggr] \; . \label{2.17}
\end{eqnarray}

The dynamic chiral susceptibility in the 1-Loop HTL approximation, considered here, has a
cubic ultraviolet divergence, which can be removed, for example, by subtracting the free result
(\ref{eq17}). We will not discuss the dynamic susceptibility further on as it is not 
directly related to QCD lattice results \cite{laer}. We will just mention here that the 
(ultraviolet divergent) dynamic
susceptibility, (\ref{2.15}) to (\ref{2.17}), normalized to the free case (\ref{eq17})
is equal to unity, as can be seen from the scalar meson correlator at $\tau = \beta$
discussed in Ref.\cite{karsch}.

\section{Conclusions}

In the present paper we have considered the chiral susceptibility of the quark-gluon
plasma within lowest order perturbation theory (free case) and the HTL approximation.
We have shown that there are two possible definitions of the chiral susceptibility
leading to different results due to a missing conservation law associated with the
chiral susceptibility in contrast to the quark number susceptibility \cite{mun}.
Using the definition based on the derivative of the chiral condensate with respect to the current 
quark mass, we have computed this static chiral susceptibility within perturbative QCD and
using an effective
HTL resummed quark propagator. In the case of the usual perturbation theory we found that
the chiral susceptibility cannot be computed beyond the leading order (free case) as it diverges
due to setting the bare quark mass equal to zero at the end. 
Similar as in the case of the free energy \cite{andersen00}
but in contrast to the quark number susceptibility \cite{mun}, the chiral susceptibility 
using the HTL approach is
ultraviolet divergent. Even after subtracting the free susceptibility or the
zero temperature part, a logarithmic temperature dependent singularity remains as
in the case of the gluonic part of the free energy \cite{andersen00}. 
Applying dimensional
regularization the HTL chiral susceptibility can be larger or smaller
than the free one depending on the choice of the renormalization constant $\mu$. Here we
expanded the chiral susceptibility up to order $m_q^4$. The $m_q^2$ contribution is the
lowest order correction to the free susceptibility, which has no $m=0$ divergence. 
The temperature dependence of the HTL chiral susceptibility 
is entirely determined by the temperature dependence of the running coupling 
constant that enters through the medium effects (quasiparticle quark mass, Landau damping)
contained in the HTL approximation.
No indication of a peak at the critical temperature is found, as seen 
in lattice QCD simulations \cite{laer}. This indicates that this peak is due to 
truly non-perturbative effects such as chiral restoration, not included in the
HTL approach.

\bigskip

{\bf Acknowledgements:}

\medskip

We would like to thank T. Kunihiro, F. Karsch and, in particular, 
M. Strickland for useful correspondence. M.G.M. is also 
thankful to E. Braaten, C. Greiner, K. Gallmeister, S. Juchem, and A. Peshier 
for various discussion. M.H.T. was supported by the DLR (BMBF) under grant no. 50WM0038. 
M.G.M. gratefully acknowledges the support from AvH Foundation and from  GSI (Darmstadt).

\vspace{0.3in}

\appendix
\renewcommand{\theequation}{\thesection.\arabic{equation}}
\section{}

\subsection{One-loop sum-integrals}
The one-loop fermionic sum-integrals necessary for our purpose in (\ref{8})
are:
\begin{equation}
\SI \frac{1}{K^2} = \left ( \frac{\mu}{4\pi T}\right )^{2 \epsilon}
\frac{T^2}{24} \left [ 1 + \left (
2 \gamma + 2\log \pi -2 \frac{\zeta^\prime(2)}{\zeta(2)}\right ) \epsilon
+{\cal O}(\epsilon^2) \right ] \ . \label{a1}
\end{equation}

\begin{equation}
\SI \frac{1}{k^4K^2}=\SI\frac{k_0^2}{k^6K^2} 
= \left ( \frac{\mu}{4\pi T}\right )^{2 \epsilon}
\frac{7\zeta(3)}{(2\pi)^4 T^2} \left [ 1 + \left (
2 + \frac{8}{7}\log 2 +2 \frac{\zeta^\prime(3)}{\zeta(3)}\right )
\epsilon +{\cal O}(\epsilon^2) \right ]  \ . \label{a2}
\end{equation}

\begin{equation}
\SI \frac{1}{(K^2)^2} = \frac{1}{(4\pi)^2} 
\left ( \frac{\mu}{4\pi T}\right )^{2 \epsilon}
\left [ \frac{1}{\epsilon} + 2 (\gamma+2\log 2)
+ \left ( \frac{\pi^2}{4} +4\gamma \log 2 + 4 \log^2 2 + 4 \gamma_1 \right )
\epsilon + {\cal O}(\epsilon^2) \right ]  \ . \label{a3}
\end{equation}

\begin{equation}
\SI \frac{1}{k^2K^4}
= \left ( \frac{\mu}{4\pi T}\right )^{2 \epsilon}
\frac{7\zeta(3)}{2(2\pi)^4 T^2} \left [ 1 +  \left (
4 + \frac{16}{7}\log 2 +2 \frac{\zeta^\prime(3)}{\zeta(3)}\right )
\epsilon +{\cal O}(\epsilon^2) \right ]  \ . \label{a4}
\end{equation}

\begin{equation}
\SI \frac{1}{(K^2)^3} = \frac{21 \zeta(3)}{128 \pi^4 T^2} 
\left ( \frac{\mu}{4\pi T}\right )^{2 \epsilon}
\left [1+ \left ( \frac{2}{3} +\frac{16}{7}\log 2-\log \pi +
2  \frac{\zeta^\prime(3)}{\zeta(3)} \right )\epsilon
+{\cal O}(\epsilon^2) \right ] \ . \label{a5}
\end{equation}

All these above sum-integrals are calculated using standard contour 
integration prescription. The number $\gamma_1$ in (\ref{a3}) is the 
first Stieltjes gamma constant.

\subsection{One-loop HTL sum-integrals}
In this subsection we will compute the one-loop sum-integrals in (\ref{8})
involving the
HTL function ${\cal T}_d(K)$ using the method developed in Ref.~\cite{Andersen02a}.

The following sum-integral in (\ref{8}) can be decomposed using (\ref{1b}) as
\begin{equation}
\SI \frac{1}{k^2 K^4} {\cal T}_d(K) = \left \langle \frac{1}{1-c^2}\right 
\rangle_c \ \ \ \SI \frac{1}{k^2 K^4} - \left \langle \frac{c^2-c^{3+2\epsilon}}
{(1-c^2)^2}\right \rangle_c \ \ \ \SI \frac{1}{k^4 K^2}  \ . \label{a6}
\end{equation}
The one-loop sum-integrals are given in (\ref{a2}) and (\ref{a4}). 
The angular averages are performed  using (\ref{1b}), leading to

\begin{equation}
\left \langle \frac{1}{1-c^2}\right \rangle_c \ \
 = -\frac{1}{2\epsilon}+1 , 
\label{a7}
\end{equation}

\begin{equation}
 \left \langle \frac{c^2-c^{3+2\epsilon}}
{(1-c^2)^2}\right \rangle_c \ \ = -\frac{1}{4\epsilon} + \left (\frac{1}{4} 
-\log 2 \right ) + \left(\frac{3}{4}-\frac{\pi^2}{6}+2\log 2-\log^2 2\right )
\epsilon + {\cal O}(\epsilon^2) . \label{a8}
\end{equation}
Once these expressions are combined, the pole in $1/\epsilon$ cancels and we obtain
a finite result:
\begin{equation}
\SI \frac{1}{k^2 K^4} {\cal T}_d(K) = \frac{7\zeta(3)}{4 (2\pi)^4 T^2}
\Big [ \frac{20}{7} \log 2 -1 \Big ] \ \ . \label{a9}
\end{equation}

Following the same strategy, the remaining sum-integral in (\ref{8}) 
involving ${\cal T}_d^2(K)$ can be written as
\begin{equation}
\SI \frac{1} {k^2k_0^2K^2} {\cal T}_d^2(K) = \left \langle 
\frac{-c_1^3 + c_2^2 (-1+c_1^3) + c_2^3 -c_1^2 (-1+c_2^3)}
{(c_1^2-c_2^2) (-1+c_1^2) (-1+c_2^2) } \right \rangle_{c_1,c_2} \ \ \ 
\SI \frac{k_0^2}{k^6 K^2} \ .
\label{a10}
\end{equation}
where $\epsilon$ has  been set to zero, as this integral is finite. The result
is given as
\begin{equation}
\SI \frac{1} {k^2k_0^2K^2} {\cal T}_d^2(K) = \frac{7 \zeta(3)}{ (2\pi)^4 T^2} 
\left [ 2\log 2 -\frac{\pi^2}{12} \right ] \ \ . \label{a11}
\end{equation}



\begin{thebibliography}{1}
\bibitem{satz}S. Digal, E. Laermann, and H. Satz, Euro. Phys. C {\bf 18}, 583 
(2001).
\bibitem{hatsuda} T. Hatsuda and T. Kunihiro, Phys. Rep. {\bf 247}, 221 (1994);
T. Kunihiro, Phys. Lett. B{\bf 271}, 395 (1991).
\bibitem{laer} F. Karsch and E. Laermann, Phys. Rev. D {\bf 50}, 6954 (1994).
\bibitem{smilga} A. Smigla and J.J.M. Verbaarschot, Phys. Rev. D {\bf 54}, 1087
(1996).
\bibitem{zhuang} P. Zhuang, J. H\"ufner, and S.P. Klevansky, Nucl. Phys. A {\bf 576}, 525
(1994).
\bibitem{blaschke} D. Blaschke, A. Hoell, C.D. Roberts, and S.M. Schmidt, Phys. Rev. 
C {\bf 58}, 1758 (1998). 
\bibitem{mun} P. Chakraborty, M.G. Mustafa, and M.H. Thoma, 
Euro. Phys. J. C {\bf 23}, 591 (2002).
\bibitem{Andersen99} J.O. Andersen, E. Braaten, and M. Strickland,
Phys. Rev. Lett. {\bf 83}, 2139 (1999).
\bibitem{andersen00} J.O. Andersen, E, Braaten, and M. Strickland, Phys. Rev. D {\bf 61},
014017 (1999); {\it ibid} 074016 (2000).
\bibitem{htl} E. Braaten and R.D. Pisarski, Nucl. Phys. B {\bf 337}, 569 (1990).  
\bibitem{kapus} J.I. Kapusta, {\it Finite-Temperature Field Theory} (Cambridge
University Press, Cambridge, 1989).
\bibitem{mclerran} L. McLerran, Phys. Rev. D {\bf 36}, 3291 (1987).
\bibitem{muk} K. Mukherjee, {\it hep-ph/0106236}.
\bibitem{KLS} J.I. Lapusta, P. Lichard, and D. Seibert, Phys. Rev. D {\bf 44}, 2774 (1991).
\bibitem{fluc} H.B. Callen and T.A. Welton, Phys. Rev. {\bf 122}, 34 (1961);
R. Kubo, J. Phys. Soc. Japan {\bf 12}, 570 (1957); D. Forster, {\it Hydrodynamic
fluctuation, broken symmetry and correlation functions} (Benjamin/Cummings,
Melno Park, 1975).
\bibitem{karsch} F. Karsch, M.G. Mustafa, and M.H. Thoma, Phys. Lett. B
{\bf 497}, 259 (2001).
\bibitem{blaizot} J.-P. Blaizot, E. Iancu, and A. Rebhan, {\it hep-ph/0206280}.
\bibitem{weld} H. A. Weldon, Phys. Rev. D {\bf 26}, 2789 (1982).
\bibitem{bpy}E. Braaten, R.D. Pisarski, and T.C. Yuan, Phys. Rev. Lett. {\bf 64}, 2241
(1990).
\bibitem{Blaizot01} J.-P. Blaizot, E. Iancu, A. Rebhan, Phys. Rev. D {\bf 63}, 065003 (2001).
\bibitem{Andersen02} J.O. Andersen and M. Strickland, Phys. Rev. D {\bf 66}, 105001 (2002).
\bibitem{Andersen02a} J.O. Andersen, E. Braaten, E. Petitgirard, and M. Strickland, 
Phys. Rev. D {\bf 66}, 085016 (2002).
\bibitem{private} F. Karsch, private communications.

\end{thebibliography}
\end{document}